\documentclass[sigconf]{acmart}

\AtBeginDocument{%
  }

\setcopyright{acmcopyright}
\copyrightyear{2022}
\acmYear{2022}
\acmDOI{XXXXXXX.XXXXXXX}

\acmConference[]{}{}{}
\acmPrice{}
\acmISBN{}

\usepackage[font=small,labelfont=bf]{caption}
\begin{document}

\title{CaminAR: Supporting Walk-and-talk Experiences for Remote Dyads using Augmented Reality on Smart Glasses }

\author{Victor Chu}

\author{Andrés Monroy-Hernández}
\affiliation{
  \institution{Princeton University}
  \city{Princeton}
  \state{New Jersey}
  \country{USA}
}

\renewcommand{\shortauthors}{Chu and Monroy-Hernández}
\begin{abstract}
  In this paper, we propose CaminAR, an augmented reality (AR) system for remote social walking among dyads. CaminAR enables two people who are physically away from one another to synchronously see and hear each other while going on a walk. The system shows a partner's avatar superimposed onto the physical world using smart glasses while walking and talking. Using a combination of visual and auditory augments, CaminAR simulates the experience of co-located walking when dyads are apart.
\end{abstract}




\maketitle

\section{Introduction}
In this paper, we design and discuss an augmented reality remote social walking experience (CaminAR) that uses smart glasses to superimpose a virtual avatar of a remote partner onto the participant's environment. By adding a virtual avatar in front of the participant, we aimed to create a perception of closeness between them and their partner beyond audio alone.

We evaluated  how CaminAR's combination of visual and auditory augmentations influenced (1) participants' ability to have a conversation while walking, (2)  social presence, and (3) connectedness between the participant and their partner. We found that CaminAR helped participants and their partner enter a shared reality space, however participants did not like the avatar's position and unrealistic responsiveness. 

\begin{figure}
\centering
\begin{minipage}{.5\columnwidth}
  \centering
  \includegraphics[width=.9\linewidth]{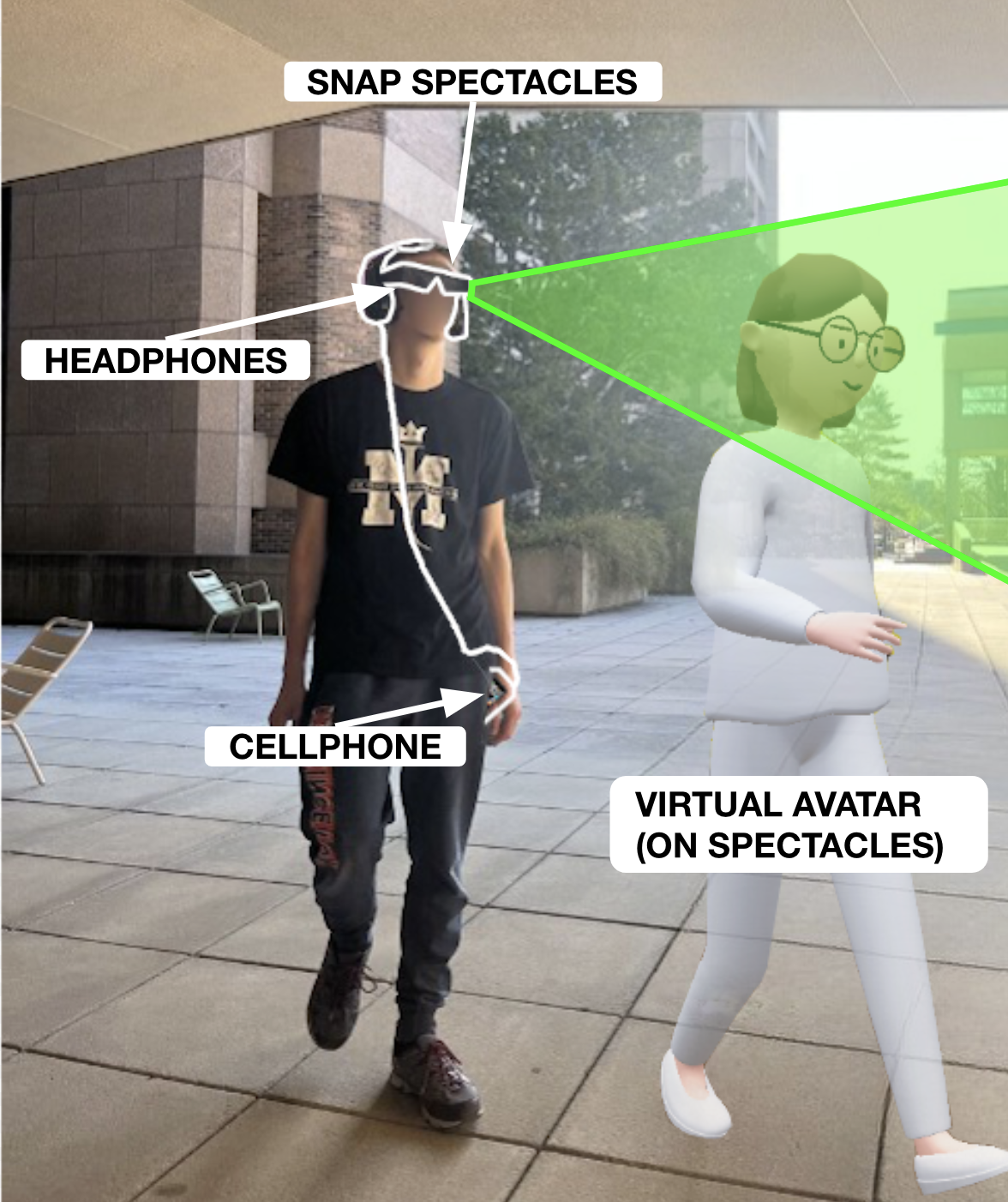}
  
  \label{fig:test1}
\end{minipage}%
\begin{minipage}{.5\columnwidth}
  \centering
  \includegraphics[width= .9\linewidth]{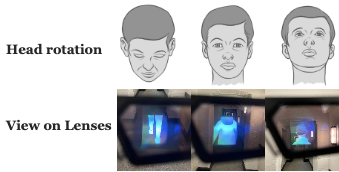}
  
  \label{fig:test2}
\end{minipage}
 \caption{Participant walking and seeing an avatar of their remote partner (left). Actual screen captures of CaminAR (right) on smart glasses based on various head rotations.}
 \bigskip
 \label{fig:snap}
\end{figure}

\section{Related Work}

Prior work on remote social exercise has explored the use of auditory and haptic augmentations. For instance, researchers created a system that enabled two participants to run at a distance \cite{supporting} while experiencing their partner's relative pace through the spatial audio. The researchers found that the system created a ``psychological sense of togetherness'' among participants \cite{design}. Similarly, other researchers investigated the use of haptic augmentations by creating a system that allowed participants to feel a remote partner's gait through vibrotactile anklets. The researchers found that participants consistently reported that the vibrotactile devices were able to effectively convey the presence of a remote companion\cite{haptic}. These two research systems show the potential for supporting remote social walking through digital augmentations.




Research on how immersion and media richness affect a participant's ability to understand another's thoughts and emotions (copresence/social presence) has shown that higher media richness results in increased levels of social presence \cite{presence}. Similarly, higher levels of immersion have been shown to evoke the state of mind of other people as if they were in the same environment \cite{immersive}. In this work, we aim to improve both the sensory fidelity of remote social walking and the immersiveness of the system by mixing multiple senses -- hearing and sight. We designed our system such that a steady presence of the virtual avatar in the participant's visual field would remind participants of the presence of their remote partner, improving the connection between the dyads\cite{connectedness}.

\section{CaminAR System}
CaminAR enables a participant using smart glasses to see a digital avatar representing a remote partner superimposed onto their environment as they go on a walk at the same time (See Fig. \ref{fig:snap}). The participant and their partner can talk using their voices while they view their partner's avatar through smart glasses. As the participant changes their direction, the partner's avatar stays in front of them. CaminAR was implemented in Lens Studio\footnote{Lens Studio is an AR authoring environment available at \url{https://ar.snap.com/}} as an AR application for the Spectacles smart glasses\footnote{Spectacles are Snap's smart glasses available at \url{https://spectacles.com}}.

Before the walk, the participant's partner selects one of six available avatars to represent them while they walk. To begin a CaminAR session, participants schedule a walk with a remote partner. When both parties are ready to walk, they call each other using earbuds.  Those with smart glasses can then wear them and start the CaminAR app. CaminAR works even if only one of the participants has smart glasses; however, only the person with smart glasses can see the visual augmentation.

\section{Evaluation}

We recruited seven university students to use CaminAR with a remote partner of their choosing. We asked participants to schedule a 20-minute social walk with someone they felt comfortable talking on the phone for that length of time. On the day of their scheduled walk, the participants were brought outside to an isolated and flat area where they would be able to roam while walking. During the walk, they spent 10 minutes with CaminAR and the other 10 minutes without it. We randomized the order of each condition. We gave a list of conversation starters, e.g., ``what is the last book you have read?'', to prompt people's conversations. We then interviewed participants to learn about their experience. See Appendix for the interview protocol.

\section{Study Results}

\subsubsection{Participants Entered a Shared Reality Space}
Four out of seven participants suggested that the visual augmentations allowed them to enter into a shared reality space with the remote partner.
One participant noted that \emph{''When there is only audio, you have to imagine a lot more of the other side of the interaction ... With [CaminAR], there is less work. You actually can see a person.''} (P3). This account illustrates the potential of visual augments to frame a space that combines the environment of both the participant and the partner during a remote social walking experience. In this shared reality, participants found that CaminAR both increases the presence of the remote partner and improves their ability to hold a conversation while walking.

\subsubsection{Avatar Position was Awkward for Participants}
Three of seven participants found that facing the avatar's back, rather than the face, made them feel disconnected. Participants wanted the experience to simulate co-located social walking where a friend would walk \textit{beside} them and not ahead, which we could not implement due to the limited field of view. That said, a participant noted that the avatar was \emph{``more of a reminder that you are talking to human-being rather than being a friend''} (P6).
Although feedback on avatar position points to areas of improvement for our implementation, a comparison of the current system to co-located social walking goes to show the potential of visual augments in this space.

\subsubsection{Participants Wanted Avatar to be More Responsive}
When we asked how the system could improve, five of seven participants noted that they wanted more varied and typical interactions between the participant and the avatar. Specifically, participants noted that \textit{``it felt weird that [the avatar] didn't turn towards [the participant]''} (P2), making it hard to see the avatar's face during the walk. In our implementation, regardless of the participant's velocity, direction, and whether they were speaking, the avatar is always in front with the back of its head facing the participant. With face-to-face interaction being considered the baseline for social presence \cite{presence}, we learned that CaminAR made participants desire higher levels of emotional connection. 


\section{Limitations}
We identified several challenges and opportunities for building AR systems for social walking, however our findings come from small and homogeneous group of participants, i.e., college students. Future work should explore recruiting a broader range of participants and relationships, e.g., co-workers, and conversations, e.g., work-related. Furthermore, all of our studies involved only one of the dyad wearing smart glasses, future work should involve both individuals wearing smart glasses.

\section{Discussion}
The results from our study teach us the tension between \textit{expectations} and \textit{immersiveness}. As immersion increases in augmented reality experiences, comparisons to real-life experiences become more natural. With these comparisons come heightened \textit{expectations} from participants. As seen in our study, participants bring up comparisons to co-located social walking that would have unlikely come up if there were only audio augmentations. While there is much room for improvement as noted by our participants, these numerous references to co-located walking and a desire for higher levels of social presence in our study are the biggest indicator of visual AR's potential for remote socialization.

To improve upon the current system, we should first explore designs where the avatar is beside the participant and outside of the field of view of the smart glasses for most of the experience.  While the avatar is ``off camera'' when the participant looks straight, the participant can turn their head to see the avatar walking alongside them. Such a modification has the potential to allow the avatar to further increase both social presence and connectedness with the remote partner compared to just audio alone. We could further incorporate spatial audio to CaminAR so that the sounds of the remote partner in the participant's headphones are oriented to come from the avatar's relative position. This way, during the call, the partner's voice comes from where the avatar is placed, inviting the participant to look at the avatar.

\bibliographystyle{ACM-Reference-Format}
\bibliography{report}

\appendix
\section{Appendix}
\subsection{Study Protocol}
Below is the process we asked participants to follow during the study:
\begin{enumerate}
\item  Each participant was asked to schedule a phone call with a remote partner during their scheduled time slot that they felt comfortable talking with for over twenty minutes. 
    
\item The participant then asks the remote partner which of the six provided avatars the partner would like to have represent them.
    
\item  During the 20 minutes, participants will spend roughly 10 minutes with CaminAR and the other 10 minutes without (the order of treatments is randomized).
    
\item  Afterwards, they will be asked the following questions in a recorded interview:
    
    \begin{itemize}
      \item What are your thoughts on the CaminAR experience? 
      \item How did your partner’s avatar’s presence change/influence your ability to have a conversation while walking?
      \item Did CaminAR impede your ability to walk around?
      \item How do you think the position of your partner’s avatar was connected to yours? How did it affect your experience?
      \item Currently, your partner’s avatar is mostly set directly in front of you. If there were no constraints, where would you want to put your partner’s avatar?
      \item How do you think the position of your partner’s avatar was connected to yours? How did it affect your experience?
      \item Currently, your partner’s avatar is mostly set directly in front of you. If there were no constraints, where would you want to put your partner’s avatar?
      \item Currently, your partner’s avatar adapts to user head-turning/direction changing by waiting for a couple hundred iterations, then moving your partner’s avatar smoothly towards the user’s view. What were your opinions on this feature? How did this feature impact your walk and talk experience?
      \item Did you notice how your partner’s avatar “adapts” to your walking speed? What were your opinions on this feature?  How did this feature impact your walk and talk experience?
      \item Are there any general suggestions for the application?
      \begin{itemize}
          \item Are there any ways you would like to interact with your partner’s avatar that you can’t as of right now?
          \item Is there anything you would do away with?
      \end{itemize}
    \end{itemize}
\end{enumerate}

\end{document}